# A Phenomenological Model for the Quantum Capacitance of Monolayer and Bilayer Graphene Devices

George S. KLIROS

Hellenic Air-Force Academy,
Department of Electronics and Communication Engineering,
Dekeleia Air-Force Base, Attica GR-1010, Greece.
E-mail: gsksma@hol.gr

**Abstract:** Graphene nanostructures exhibit an intrinsic advantage in relation to the gate delay in three-terminal devices and provide additional benefits when operate in the quantum capacitance limit. In this paper, we developed a simple model that captures the Fermi energy and temperature dependence of the quantum capacitance for monolayer and bilayer graphene devices. Quantum capacitance is calculated from the broadened density of states taking into account electron-hole puddles and possible finite lifetime of electronic states through a Gaussian broadening distribution. The obtained results are in agreement with many features recently observed in quantum capacitance measurements on both gated monolayer and bilayer graphene devices. The temperature dependence of the minimum quantum capacitance around the charge neutrality point is also investigated.

## 1. Introduction

Graphene, an atomic layer of carbon atoms arranged in a two dimensional (2D) honeycomb lattice, are highly promising candidate for new semiconductor materials and devices [1]. In monolayer form, is gapless as its conical conduction and valence bands touch at two inequivalent Dirac-points where the density of states vanishes. The key property of graphene for electronic applications is the fast electronic transport expressed by its high carrier mobility. Since monolayer graphene has no band-gap, it is not directly suitable for digital electronics, but is very promising for analog, high frequency applications [2-4].

A unique feature of both monolayer and bilayer graphene is that the density of carriers can be tuned continuously by an external gate from electron-like carriers at positive doping to hole-like at negative doping [5]. The behavior at the crossover depends on the disorder which induces regions of inhomogeneous carrier density i.e., puddles of electrons and holes [6]. Tuning the carrier density by the gate voltage, the ratio between electron puddles and hole puddles changes until at very high densities there is one type of carriers. An important difference between monolayer and bilayer graphene is the band structure near the Dirac point. Monolayer graphene has a conical band structure and a density of states that vanishes linearly at the Dirac point. Bilayer graphene has a hyperbolic band structure and a density of states rising linearly with increasing energy from a finite value at zero energy.

Bilayer graphene has attracted great interest due to the fact that an energy gap could be opened by chemical doping or by applying external perpendicular electric field. Ones could exploit this property to use bilayer grapheme as a channel material for FETs, defining an energy gap when it is really needed, i.e. when the device must be in the off state [7,8]. Moreover, bilayer graphene patterned with a periodic array of metallic gate electrodes could replace the existing semiconductor superlattices [9].

One of the main characteristics of FETs is the capacitance formed between the channel and the gate. It is well known that the capacitance in these devices is dominated by the capacitance of the oxide layer which makes difficult to extract the quantum capacitance [10]. However, in order to decrease the operating voltage, it is expected that the oxide layers will be much thinner and have higher values of dielectric constant, which means that the quantum capacitance will be the dominant source of capacitance [11]. As a consequence, quantum capacitance is important for understanding the fundamental electronic properties of the material such as the density of states as well as device performance including the I-V characteristics and the device operation frequency.

Recently, graphene sheets have been subject to theoretical as well as experimental studies of the quantum capacitance [12-15]. Measurements on the quantum capacitance of bilayer graphene have been shown similar behaviour to that of monolayer graphene but, near the Dirac point, a finite capacitance value has been found. In order to provide physical insight into the capacitance of graphene devices, it is important to develop intuitive analytical models capturing the essential physics of the device at hand. In this paper, a simple analytical model for the quantum capacitance of both monolayer and bilayer graphene devices, is presented. The model takes into account the broadening of the density of states due to electron-hole puddles induced by local potential fluctuations and possibly to finite lifetime of electronic states. The temperature dependence of quantum capacitance is also investigated.

## 2. Quantum Capacitance Modeling

Monolayer graphene is a zero-gap semiconductor because its conducting and valence π-electron bands touch each other only at two isolated points in its two-dimensional (2D) Brillouin zone. The dispersion relation of these bands in the vicinity of these points is given by [1]

$$E_s(k) = s(\hbar v_F k) \qquad (1)$$

where $s=+1$ for the conduction band (CB) and $s=-1$ for the valence band (VB), $v_F = \sqrt{3}\gamma_0 a/2$, is the Fermi velocity with intralayer coupling $\gamma_0=3.16$ eV and $k$ is the wave vector of carriers in the two-dimensional plane of the graphene sheet. The point $k=0$, referred to as the "Dirac point," is a convenient choice for the reference of energy; thus, $E(k=0)=0\ eV$.

Bilayer graphene is composed of a pair of honeycomb lattices of carbon atoms, which include $A_1$ and $B_1$ atoms on layer 1 and $A_2$ and $B_2$ on layer 2. As shown in Fig. 1(a), the two layers are arranged in Bernal stacking, where $A_2$ atoms are located directly below $B_1$ atoms. The lattice constant within a layer is given by $a=0.246$ nm and the layer spacing by $d=0.334$ nm. In the absence of disorder, the bandstructure of clean bilayer graphene can be written [16]

$$E_{\mu,s}(k) = s\left(\mu\frac{\gamma_1}{2} + \sqrt{\frac{\gamma_1^2}{4} + (\hbar v_F k)^2}\right) \qquad (2)$$

where $\mu = \pm 1$, $s = \pm 1$, $\gamma_1=0.39$ eV is the interlayer coupling. The index $\mu = (-)$ gives a pair of bands closer to zero energies, and $\mu = (+)$ another pair repelled away by approximately $\pm\gamma_1$. In each pair, $s=(+1)$ and $(-1)$ represent the electron (CB) and hole (VB) branches, respectively. Thus, as shown in Fig.1(b), the band structure of bilayer graphene is quadratic at small momenta, like a two-dimensional electron gas, and becomes linear with increasing momentum like monolayer graphene. However, recent experimental data have revealed a hyperbolic and asymmetric band structure without a constant density of states expected for a quadratic dispersion [15].

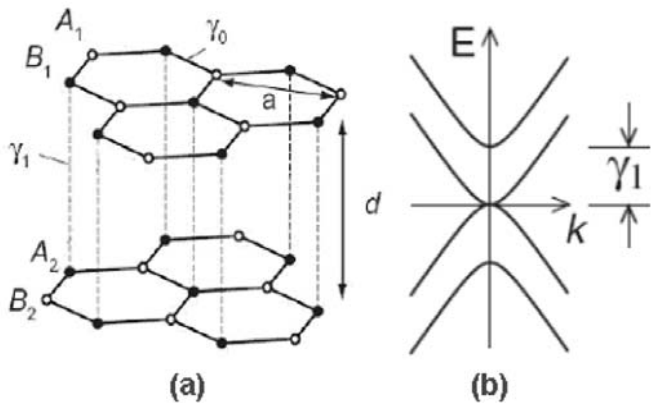

**Fig.1.** Schematic view of bilayer graphene in Bernal stacking (a) and low energy bands of perfect bilayer grapheme (b).

The density of states of pure and perfect monolayer graphene is given by

$$g_{MLG}(E) = \frac{g_s g_v}{2\pi(\hbar v_F)^2}|E| \quad (3)$$

where $g_s$, $g_v$ is the spin and valley degeneracy respectively. For the energy range $|E| \leq \gamma_1$, the density of states of the pure and perfect bilayer graphene can be well approximated by a linear relation as a function of energy [17]

$$g_{BLG}(E) = \frac{g_s g_v}{2\pi(\hbar v_F)^2}\left(|E| + \frac{\gamma_1}{2}\right) \quad (4)$$

where $g_s$, $g_v$ is the spin and valley degeneracy respectively. The above equation is accurate enough for low to moderate doping levels such that the chemical potential is less than 1 eV and is only incurs a relative error of up to a few percents when is between 1 eV and 2 eV. Near the Dirac point, the density of states is given by

$$g_{BLG}(E) = \frac{g_s g_v}{2\pi(\hbar v_F)^2}\frac{\gamma_1}{2} = g_s g_v \frac{m^*}{2\pi \hbar^2} \quad (5)$$

which is the formula for the density of states of two-dimensional electron gas with an effective mass, $m^* = \gamma_1/(2 v_F^2)$, that is, proportional to interlayer coupling.

Generic to both monolayer and bilayer graphene samples on a substrate are the so-called 'electron-hole' puddles induced by charged impurities which lead to inhomogeneous variations in the carrier density across the sample over a typical scale of tens of nm [6]. To take into account the electron-hole puddles and possible finite lifetime of electronic states, we introduce a Gaussian broadened density of states $D(E)$ as follows

$$D(E) = \frac{1}{\sqrt{2\pi}\,\Gamma}\int_{-\infty}^{+\infty}\exp\left(-\frac{(\varepsilon-E)^2}{2\Gamma^2}\right)g(\varepsilon)\,d\varepsilon \quad (6)$$

where $\Gamma$ is an energy broadening parameter which is the only phenomenological parameter of our model. After the integration in (6), we obtain

$$D_{BLG}(E) = \frac{g_s g_v}{2\pi(\hbar v_F)^2}\times\left[\frac{2\Gamma}{\sqrt{2\pi}}\exp\left(-\frac{E^2}{2\Gamma^2}\right) + E\,\text{erf}\left(\frac{E}{\Gamma\sqrt{2}}\right) + \frac{\gamma_1}{2}\right] \quad (7)$$

where $\text{erf}(x)$ is the Gaussian error function. Near the Dirac point, the density of states becomes

$$D_{BLG}(0) = g_s g_v \frac{m^*}{2\pi\hbar^2}\left(1 + 2\sqrt{\frac{2}{\pi}}\frac{\Gamma}{\gamma_1}\right) \quad (8)$$

which increases linearly with the broadening parameter $\Gamma$. For $\gamma_1 = 0$ equation (7) leads to the broadened density of states for the monolayer graphene

$$D_{MLG}(E) = \frac{g_s g_v}{2\pi(\hbar v_F)^2}\times\left[\frac{2\Gamma}{\sqrt{2\pi}}\exp\left(-\frac{E^2}{2\Gamma^2}\right) + E\,\text{erf}\left(\frac{E}{\Gamma\sqrt{2}}\right)\right] \quad (9)$$

The quantum capacitance is defined as the derivative of the total net charge of the monolayer or bilayer graphene device with respect to applied electrostatic potential. The total charge is proportional to the weighted average of the density of states at the Fermi level $E_F$. When the density of states as a function of energy is known, the quantum capacitance $C_Q$ of the channel at finite temperature can be calculated as [18]

$$C_Q = e^2 \int_{-\infty}^{+\infty} D(E)\left(-\frac{\partial f(E-E_F)}{\partial E}\right) dE \qquad (10)$$

where *f(E)* is the Fermi-Dirac distribution. The above relation is strictly valid only when the electrostatic potential is position independent.

The quantum capacitance describes the response of the charge inside the graphene-channel to the conduction and valence band movement and is a strong function of Fermi energy $E_F$ which can be changed experimentally by the gate voltage $V_G$. This distinguishes graphene from conventional two-dimensional electron systems in which the quantum capacitance is usually a small and constant contribution that is difficult to be extracted from the experimental data. In the following section we present our numerical results for the quantum capacitance of both monolayer and bilayer graphene devices based on the Eqs. (7), (9) and (10).

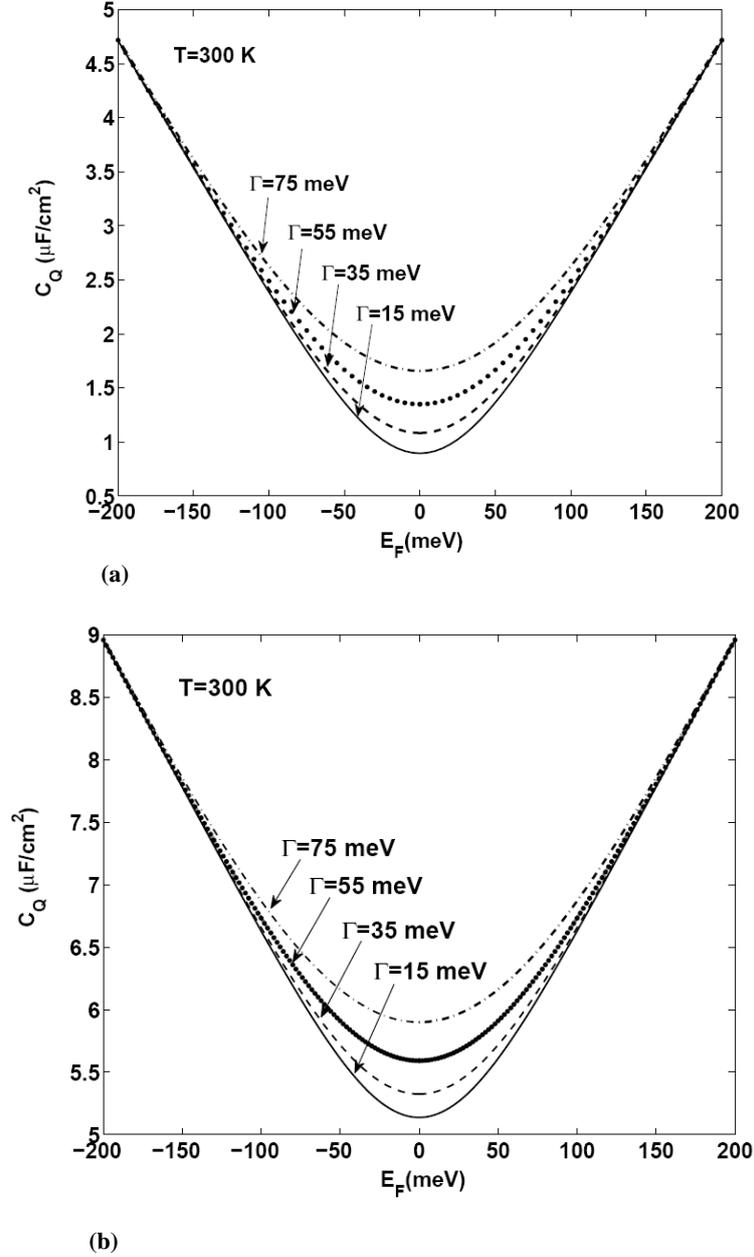

**Fig.2.** Calculated quantum capacitance versus Fermi energy for broadening parameters
Γ=15, 35, 55 and 75 meV in monolayer (a) and bilayer (b) graphene device.

## 2. Results and discussion

Fig. 2 shows the quantum capacitance of monolayer (a) and bilayer (b) graphene as a function of Fermi energy at room temperature 300 K, for different broadening parameters Γ. This energy broadening range corresponds to a carrier density variation $\delta n \sim 3.5 \times 10^{11}$ cm$^{-2}$ which is consistent with the literature value attributed to electron-hole puddles in graphene on SiO$_2$ [19].

Several important features are worth noting which are in good agreement with recent experimental results, but we do not make an attempt to obtain quantitative agreement since the experimental results show substantial sample-to-sample variation. Instead we discuss the qualitative features of our numerical results: First, the quantum capacitance has a minimum value at the Dirac point which increases with the broadening parameter Γ. Second, the capacitance minimum regime becomes increasingly round and far from this regime the capacitance becomes linear with decreasing slope as Γ increases. Finally, the capacitance curve is symmetric with respect to the Dirac point.

The temperature dependence of the quantum capacitance is shown in Fig. 3 for monolayer (a) and bilayer (b) graphene, where a value for $\Gamma$=35 meV is adopted. The minimum of quantum capacitance is round and increases as temperature increases. For Fermi energies $E_F \gg \Gamma$, the capacitance becomes approximately temperature independent. It is worth noting that the quantum capacitance of pure and perfect bilayer graphene, at very low temperatures, has finite value of about 4.3 μF/cm$^2$.

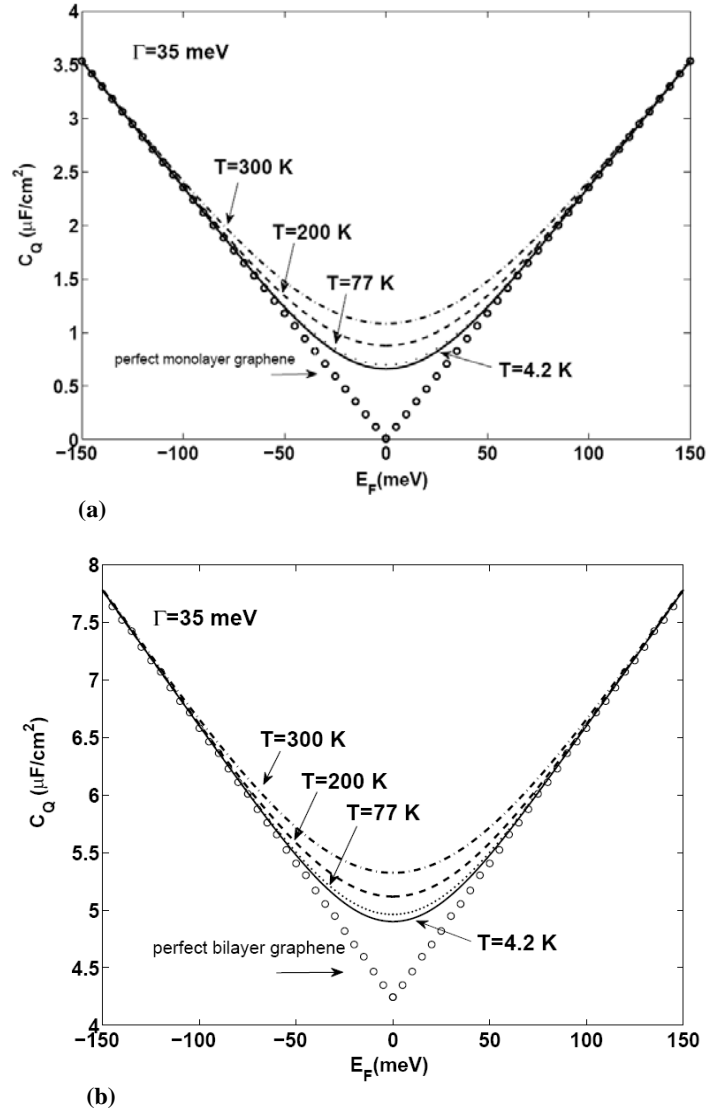

**Fig.3.** Temperature dependence of the simulated quantum capacitance versus Fermi energy for monolayer (a) and bilayer (b) graphene device.

Understanding the temperature dependence of the minimum quantum capacitance is complicated due to the activation of carriers at finite temperatures as well as to the formation of electron-hole puddles. Fig. 4 shows the minimum quantum capacitance scaled by the zero-temperature minimum capacitance as a function of temperature. As temperature increases, we observe an enhanced temperature dependence of the minimum capacitance of monolayer graphene compared to that of bilayer graphene. On the other hand, in the low temperature regime it seems that the effect of carrier density fluctuations and the associated electron-hole puddle structure is very similar to both monolayer and bilayer graphene devices.

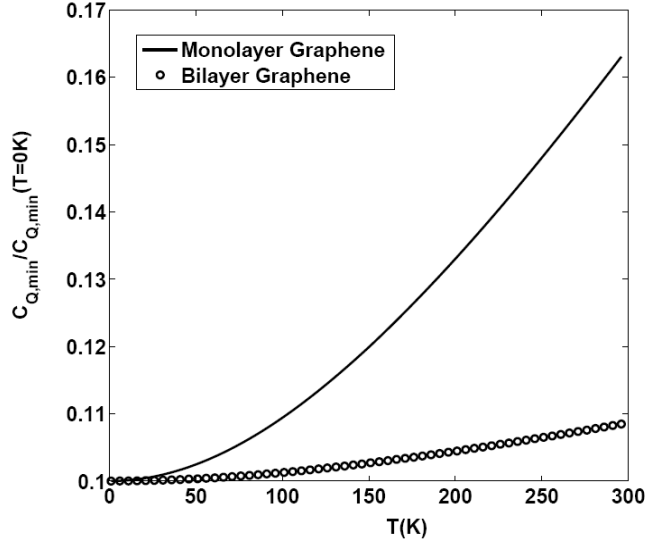

**Fig.4.** Temperature dependence of the minimum quantum capacitance of graphenes using a broadening parameter $\Gamma$ = 35 meV.

## 4. Conclusions

In conclusion, we have presented a simple phenomenological model for the quantum capacitance of both monolayer and bilayer graphene devices. Quantum capacitance is calculated from the broadened density of states taking into account electron-hole puddles and possible finite lifetime of electronic states through a Gaussian broadening distribution. Adopting a range of values for the broadening parameter $\Gamma$ between 15 meV and 75 meV, the obtained results are in agreement with many features recently observed in quantum capacitance measurements on gated nonolayer or bilayer graphene.

The quantum capacitance of both monolayer and bilayer graphene has a finite minimum value at the Dirac point which increases with the broadening parameter $\Gamma$. The minimum-value regime becomes increasingly round and far from this regime the capacitance becomes linear with decreasing slope as the energy broadening increases. The capacitance curve $C_Q(E_F)$ for monolayer and bilayer graphene devices becomes temperature independent at Fermi energies $E_F \gg \Gamma$. The temperature dependence of the minimum quantum capacitance is also studied. As temperature increases, the minimum quantum capacitance increases dramatically for bilayer graphene, while it is nearly unchanged for monolayer graphene. Moreover, in the low temperature regime it seems that the effect of the electron-hole puddle formation is very similar to both monolayer and bilayer graphene devices.

We hope that our model is a step for understanding the gate voltage and temperature dependence of the quantum capacitance of graphene devices. The phenomenological parameter $\Gamma$ can be treated as a fitting parameter and as a consequence of this parameterization, our results do not depend on the microscopic details of the impurity potential provided this parameterization describes correctly the properties of the impurity potential. However, a self-consistent effective medium theory is needed for a rigorous treatment of the broadening effects including the screening of the impurity field by the carriers.